\documentclass{amsart}

\usepackage{mathtools}
\usepackage{algorithm}
\usepackage[noend]{algpseudocode}
\usepackage{siunitx}
\usepackage[draft]{hyperref} 
\usepackage{multirow} %
\usepackage{nicefrac}
\usepackage{rotating}
\usepackage{enumerate}
\usepackage{tabu}
\usepackage[numbers, square]{natbib}
\newcommand {\xx}[1]{}
\newcommand{\overW}{\ensuremath{\overline{W}}}
\newcommand{\hatN}{\ensuremath{\hat{N}}}
\newcommand{\Var}{\ensuremath{\text{Var}}}

\numberwithin{equation}{section}

\begin{document}
\title{On estimating the alphabet size of a discrete random source.}
\author{Philip Ginzboorg}
\address{Huawei Technologies and 
              Aalto University \\  
              School of Electrical Engineering \\
              Department of Communications and Networking, Finland.}
\curraddr{Huawei Technologies, It{\"a}merenkatu 9, 00180 Helsinki, Finland}
\email{philip.ginzboorg@iki.fi}  
\thanks{}



\dedicatory{}

\begin{abstract}
We are concerned with estimating alphabet size $N$ from a stream of symbols taken uniformly at random from that alphabet. 
We define and analyze a memory-restricted variant of an algorithm that have been earlier proposed for this purpose. The alphabet size $N$ can be estimated in $O(\sqrt{N})$ time and space by the memory-restricted variant of this algorithm.
\end{abstract}

\maketitle

\section{Introduction}
\label{intro}
A discrete random source picks symbols uniformly at random from a finite alphabet of size $N$; any one of the symbols is picked with the same probability $1/N$. In this paper we investigate a memory-restricted variant of a simple algorithm that estimates $N$ from the stream of symbols that are emitted by such source. 

This algorithm has been introduced by Brassard and Bratley \cite{1988Brassard}  for finding cardinality of a set; and further analysed by  Flajolet \cite{2004Flajolet}, and Flajolet and Sedgewick \cite{2009FlajoletSedgewick}. \xx{(Their analysis includes an answer to the following variant of the Birthday Problem: "On average how many people need to be admitted into a room before we see a first birthday coincidence?" [Klamkin, M. S. and Newman, D. J. "Extensions of the Birthday Surprise." J. Combin. Th. 3, 279-282, 1967.])} Using this algorithm to estimate alphabet size of discrete random source has been proposed by Montalv\~ao et al.~\cite{2012Montalvao}.

The working of this algorithm is based on the same phenomenon as the Birthday Problem \cite{WikipediaBirthdayProblem}; it needs $O(\sqrt{N})$ time (number of observed symbols) on the average to estimate the alphabet size, which may be an advantage when $N$ is large. The space (number of stored symbols) needed for making an estimate is $O(\sqrt{N})$ on the average; but in the worst case the space needed will be $N +1$ symbols and the runtime will be $N+1$ multiplied by a constant. 

The main contribution of the present paper is in sections \ref{sec:alg,c} and  \ref{sec:error}, where we define and analyse the behavior of this algorithm when its internal memory is limited to at most $c < N$ symbols.
   
We have also included original derivation of known results for the case when there is no such restriction. This is done for ease of reference, and to show how to analyse this algorithm using relatively simple means.

In the next section we explain how the algorithm works.  Section \ref{sec:experiments} describes the results of experiments with pseudorandom number generator as the source of symbols. In sections \ref{sec:alg,c} and \ref{sec:error} we define a memory-restricted variant of the algorithm and then consider the effects of limited memory on accuracy of estimation. Section \ref{sec:theory} contains theoretical calculations.  The paper ends with conclusion in section \ref{sec:conclusion}.

\section{The algorithm}
\label{sec:estimator}
The stream of symbols coming from a discrete random source is divided on-line into adjacent, variable-sized blocks of symbols, such that one block includes a single pair of identical (matching) symbols. The sizes of these blocks comprise a sequence of random variables $W[1]$, $W[2]$,\dots. 
An example using uppercase letters as symbols will clarify this. Suppose that the following sequence of symbols, coming from a discrete random source, are observed: A, B, K, D, E, I, M, D, A,  D, C, K, A, C, J, I, \dots

The first repeating symbol in that sequence is D, and the size of the block (A, B, K, D, E, I, M, D) that contains the first pair of matching symbols D is eight. Therefore, we record $W[1] = 8$, discard the beginning of the sequence up to and including the second occurrence of D and continue scanning. The continuation of the sequence is A, D, C, K, A, C, J, I, \dots.

Now the first repeating symbol in the sequence is A; we have a block (A, D, C, K, A) of size five with two identical symbols A. Therefore, we record $W[2] = 5$, discard the beginning of the sequence up to and including the second occurrence of A and again continue scanning.
In this manner we get a series of block sizes $W[1]$, $W[2]$, $W[3]$, and so on.  

Sample mean $\overW_l$ from $l$ realizations of $W$ is 
\begin{equation*}
\overW_l = \frac{W[1] + W[2] +\dots+ W[l]}{l}.
\end{equation*}
Expected value and variance of $\overline{W}_l$ are 
$E(W)$ and 
$\Var(W)/l$, respectively. 
%
The estimation of alphabet size $N$ from $\overW_l$ in the algorithm is based on those statistics. 
We remark that measurement of $\overW_l$ may tolerate partial loss of symbols in the stream. For instance, if we delete, say, every tenth symbol from the stream, the remaining symbols will still (i) include all of the alphabet, and (ii) be uniformly distributed in the resulting sequence. As a consequence, statistics of $\overW_l$
will remain the same after that deletion.

In section \ref{sec:theory} below it is shown that theoretical mean and  variance of block size $W$ are: 
\begin{align}
\begin{aligned}
E(W) &\approx \sqrt{\frac{\pi N}{2}} + \frac{2}{3},  \qquad \text{and}
\label{eq:Knuth simplified}
\end{aligned}
\end{align}
\begin{equation}
\Var (W)  = 2N +E(W) -E(W)^2.
\label{eq:Var(W) simplified}
\end{equation}
When $N$ is large, variance is approximately
\begin{equation}
\Var(W) \approx 2N -\frac{\pi}{2}N - \frac{4}{3}\sqrt{\frac{\pi}{2}N} \qquad \text{(large $N$)}.
\label{eq:var W, large N simplifed}
\end{equation}

These expressions are the answer to an extension of Birthday Problem \cite{WikipediaBirthdayProblem}:  ``What are (1) the average size, and (2) the variance, of a group of people where exactly two group members share the same birthday?'' 

An estimator $\hatN$ of alphabet size from the sample mean, which is based on statistical method of moments, is obtained as follows: We invert \eqref{eq:Knuth simplified}, so that $N$ is expressed as a function $g$ at  point $E(W)$, 
\begin{equation}
N \approx g(E(W)) = \frac{2}{\pi}\left(E(W)-\frac{2}{3}\right)^2.
\label{eq:N from Q(N)}
\end{equation}  
From a first-order Taylor series expansion of $g$ around the point $E(W)$, it can be shown that  $g(E(W))$ approximately equals to the average value of $g(\overline{W}_l)$ \cite{2002Casella}. We apply this approximation and replace $E(W)$ in \eqref{eq:N from Q(N)} by sample mean $\overW_l$ from $l$ realizations of $W$: 
\begin{equation}
g\left(\overW_l\right)= \frac{2}{\pi}\left(\overW_l -\frac{2}{3}\right)^2.
\label{eq:variable W g}
\end{equation}  

Since $N$ is an integer, we will round down \eqref{eq:variable W g} with the floor function, resulting in estimator $\hatN$:
\begin{equation}
\hatN = \left\lfloor \frac{2}{\pi}\left(\overW_l -\frac{2}{3}\right)^2 \right\rfloor.
\label{eq:variable W}
\end{equation}

The average time (number of observed symbols) needed to make an estimate is the mean of the sum $W[1]+W[2]+\cdots+W[l]$, i.e.~the average time is $l \cdot E(W).$
Since mean block size $E(W)$ is $O(\sqrt{N})$ by equation \eqref{eq:Knuth simplified}, the average time needed to make an estimate by this algorithm is $O(\sqrt{N})$. The space required to estimate $N$ by this algorithm will be also $O(\sqrt{N})$ on the average. 

\xx{Montalv\~ao, Silva and Attux \cite{2012Montalvao} proposed a variant of this algorithm. They noticed that $N$ is approximately a quadratic function of $\overW_l$,
and found its coefficients by fitting the quadratic to simulation results using the method of least squares. The outcome is
\begin{equation}
g\left(\overW_l\right)  \approx a \overW_l^2 - b \overW_l + c,
\label{eq:E2m g}
\end{equation}
where $a=0.6366$, $b=0.8493$, and $c=0.1272$. After multiplying the terms in the theoretically-derived quadratic function \eqref{eq:variable W g} we see that its coefficients are $a=0.6366$, $b=0.8488$, $c=0.2829$. The quadratic function \eqref{eq:E2m g} has the same $a$, but different $b$ and $c$. Rounding down \eqref{eq:E2m g} with the floor function, results in the estimator $\hatN$:
\begin{equation}
\hatN = \left\lfloor a \overW_l^2 - b \overW_l +  c \right\rfloor,
\label{eq:E2m}
\end{equation}  
where $a=0.6366$, $b=0.8493$, and $c=0.1272$.
}

A theoretical calculation in section \ref{sec:E2 bias} shows that  
the result of estimating $N$ by equation \eqref{eq:variable W g} 
will have a positive bias $\alpha>0$: $E(\hatN) = N(1+\alpha)$. This is why we round down -- rather than up -- to obtain an integer in \eqref{eq:variable W}. 

For large $N$, $\alpha$ is approximately $0.27/l$; and a more accurate estimator for large $N$ can be obtained if we divide the right hand side of equation \eqref{eq:variable W g} by $(1+\alpha)$:
\begin{equation}
\hatN = \left\lfloor\frac{1}{1+\frac{0.27}{l}} \cdot \frac{2}{\pi}\cdot \left(\overW_l -\frac{2}{3}\right)^2\right\rfloor \qquad \text{(large $N$)}.
\label{eq:method 2 heuristic}
\end{equation}

Performance of estimators \eqref{eq:variable W} 
and \eqref{eq:method 2 heuristic} is compared in next section.

We will use coefficient of variation $\text{CV}=\sqrt{\Var(\hatN)}/E(\hatN)$ to characterize the extent of variability of $\hatN$ in relation to its mean. The smaller this coefficient, the more precise is $\hatN$. 
First-order approximation to the squared $\text{CV}$ is
\begin{align}
\begin{aligned}
\text{(CV)}^2 
&\approx \frac{1}{l}\cdot \frac{8}{\pi} \cdot \left(2-\frac{E(W)\cdot \left(E(W)-1\right)}{N}\right).
\end{aligned}
\label{eq:C^2 W result} 
\end{align}
%
For large alphabet sizes it reduces to
\begin{align}
\begin{aligned}
(\text{CV})^2 & \approx \frac{1.09}{l} \qquad \text{(large $N$)}.
\end{aligned}
\label{eq:C^2 W)large N} 
\end{align}




Inverting \eqref{eq:C^2 W)large N} we get the number of blocks $l$ that would be needed to estimate a large $N$ with a given coefficient of variation:
\begin{align}
\begin{aligned}
l  & \approx \left\lceil \frac{1.09}{(\text{CV})^2} \right\rceil \qquad \text{(large $N$)}.
\end{aligned}
\label{eq:l from CV large N} 
\end{align}
For example, if we want a 10 percent CV, then we would need to observe $109$ blocks, and 
this value of $l$ multiplied by average block size $E(W)$ of $\sqrt{\pi N /2} +2/3$  predicts average measurement time of about $136.6\sqrt{N} +72.7$ symbols; for a less precise measurement where target CV is 15 percent we would need to observe $49$ blocks, resulting in average measurement time of about $61.4\sqrt{N} +32.7$ symbols; and for a more precise measurement where target CV is 5 percent we would need to observe $446$ blocks, resulting in average measurement time of about $546.4\sqrt{N} +290.7$ symbols.

If distribution of symbols in the stream is non-uniform, the above algorithm will tend to underestimate $N$,  because when some symbols occur in the stream more often than others, the average block size $\overW_l$ tends to be smaller than when all symbols are equally likely to occur. Brassard and Bratley \cite{1988Brassard} note that in this case the algorithm may still be used to probabilistically estimate a lower bound on $N$. 

\section{Experiments}
\label{sec:experiments}
The working of the algorithm with estimators \eqref{eq:variable W} 
and \eqref{eq:method 2 heuristic}, and $l$ set by equation~\eqref{eq:l from CV large N} based on target CV, was tried with six alphabet sizes  $N=10$,  100,\dots, $10^6$. In each of these six experiments $N$ was held constant and symbols came from a pseudorandom sequence. This sequence was generated by Matlab's \texttt{randi} function, which, by default, uses Mersenne Twister method for producing pseudorandom numbers at the time of this writing. Target coefficient of variation in these measurements was set to 10 percent, resulting in $l=109$ by equation~\eqref{eq:l from CV large N}. For each value of $N$ both estimators computed their $\hatN$  after 109 block sizes have been collected, and this operation was repeated twenty thousand times.

Empirical bias and coefficient of variation were computed from resulting set of estimates. They are listed as percentage points in Table~\ref{table:bias and CV}. Empirical bias is the relative error between the mean of twenty thousand values of $\hatN$ and actual alphabet size $N$: $\text{bias} = (\text{mean}(\hatN) - N)/N$; empirical coefficient of variation CV is the standard deviation of twenty thousand values of $\hatN$ divided by their mean. 
\begin{table}[ht]
\caption{Empirical bias and CV as percentage points for two estimators; target CV was set to 10 \%.}
\label{table:bias and CV}
%
{\tabulinesep=1.2mm
\begin{tabu}{|c|c|cccccc|}
\hline 
\multicolumn{1}{|c}{} &  & \multicolumn{6}{c|}{$N$}\tabularnewline
\hline 
Estimation by &  & 10 & $100$ & $10^{3}$ & $10^{4}$ & $10^{5}$ & $10^{6}$\tabularnewline
\hline 
 & CV \% & $9.28$ & $9.51$ & $9.77$ & $9.88$ & $9.98$ & $9.99$\tabularnewline
\hline 
Eq.~\eqref{eq:variable W} & bias \% & $-3.29$ & $-0.03$ & $0.20$ & $0.20$ & $0.25$ & $0.27$\tabularnewline
 & CV \% & $9.21$ & $9.50$ & $9.76$ & $9.87$ & $9.98$ & $9.99$\tabularnewline
\hline 
Eq.~\eqref{eq:method 2 heuristic}  & bias \% & $-3.75$ & $-0.27$ & $-0.05$ & $-0.05$ & $-0.00$ & $-0.02$\tabularnewline
 & CV \%  & $9.23$ & $9.49$ & $9.76$ & $9.87$ & $9.98$ & $9.99$\tabularnewline
\hline 
\end{tabu}
}
\end{table}

Looking at Table~\ref{table:bias and CV} observe, first, that  in all cases the empirical coefficient of variation is below the target 10 percent. 

Next, compare the results of estimation by equations \eqref{eq:variable W} and  \eqref{eq:method 2 heuristic}. There are no significant differences in precision between the two estimators. Regarding accuracy, when $N$ is 10 and 100, the estimation by equation \eqref{eq:method 2 heuristic} is less accurate than by equation \eqref{eq:variable W}. But \eqref{eq:method 2 heuristic} is more accurate when $N$ is large, in the range $10^3$, $10^4$, $10^5$, and $10^6$. Observe that in that range the bias in estimating with \eqref{eq:method 2 heuristic} is reduced by 0.25 percent compared to estimating with \eqref{eq:variable W}. This is because   when $l=109$, the first term in equation \eqref{eq:method 2 heuristic} is approximately $1-0.25 \cdot 10^{-2}$.

All in all, estimation by equation \eqref{eq:method 2 heuristic} is best for measuring alphabet sizes in the order of $10^3$ and higher.

\section{Memory-restricted algorithm}
\label{sec:alg,c}
From equation \eqref{eq:Var(W) simplified}
we see that mean block size $E(W)$ of $\sqrt{\frac{\pi N}{2}}$, or about $1.25\sqrt{N}$ symbols is sufficient to estimate $N$. But the observed block size instances $W[k]$ will sometimes exceed $E(W)$ and grow, until $N+1$ in the worst	 case. 

On the one hand, we need to store up to $N$ previously observed symbols in computer memory in order to accurately measure all $W[k]$. 
On the other hand, the computer's memory available for this purpose may be limited to hold less than $N$ symbols when $N$ is large. Let us denote this limit on the number of stored symbols by $c$. 

If  $N$ is bigger than our storage capacity $c$, then we cannot accurately measure block sizes $W[k]$ that are greater than $c$. 
As a result our estimates of alphabet size will be smaller than $N$. 

When $N>c$, it is sensible to clip the data. This means that any block size that is greater than $c$ is replaced by $c+1$ during measurement. It is also sensible to report the number of times $Y$ that the event $W[k]>c$ has happened during measurement: if $Y$ is zero, then memory limit $c$ did not impact the measurement; and the ratio between $Y$ and $l$ can serve as a rough approximation to the probability of $W>c$. We will mark clipped block sizes by $W_c$:
$$
W_c =  \begin{cases} 
       W, &\mbox{if } W \leq c; \\
       $c + 1$, & \mbox{otherwise.}
       \end{cases}  
$$

Listing Algorithm \ref{alg:E2 w, c} shows pseudocode of resulting memory-restricted algorithm. We let the identation separate code blocks -- there are no {\bf end} statements at the end of {\bf function} definitions, {\bf if} conditions and {\bf for} loops.

Algorithm \ref{alg:E2 w, c} includes two functions:  the first, {\sc getBlockSize}, repeatedly calls {\sc getSymbol} to read a symbol from stream emittied by a discrete random source, until a matching pair of symbols is found or the memory limit $c$ is reached; it then returns the clipped block size $W_c$. 
Please note that {\sc getBlockSize} uses a table, denoted by $T$, to store unique symbols that have been observed so far in a block. Initially, before we start reading symbols into the block, table $T$ is empty. Data structure used for holding $T$ in computer memory can be a hash table keyed by observed symbols.

Recall that sample size $l$ that is needed to obtain target precision CV can be computed by equation \eqref{eq:l from CV large N}. The second function, {\sc EstimateN}, calls {\sc getBlockSize} $l$ times to obtain sample mean of $l$ block sizes; it then  computes the estimate of $N$ by equation \eqref{eq:method 2 heuristic}.

\begin{algorithm}
\caption{Measuring large $N$ given a sample size $l$ and memory limit $c$.}
\label{alg:E2 w, c}
\begin{algorithmic}
\Function{getBlockSize}{$c$}
   \State $T \gets \emptyset$ \Comment initialize table of observed symbols
   \For {$j=1:c$}
     \State $s \gets \text{\sc getSymbol}$ \Comment get symbol from the stream
      \If {$s \in T$} break \Comment exit the for loop 
      \EndIf
       \State {$T \gets T \cup s$} \Comment insert $s$ into the table 
     \EndFor
     \If {$s \notin T$} 
       {$j = c + 1$} \Comment we have hit the memory limit $c$ 
      \EndIf
\xx{\While {$s \notin T$ and $j < c $} 
\State {$T \gets T \cup s$} 
\State $s \gets \text{\sc getSymbol}$ \Comment get the next symbol of the block
\State {$j \gets j+1$}
\EndWhile
\If {$s \notin T$} 
{$j = c + 1$} \Comment we have hit the memory limit $c$ 
\EndIf
}
\State \Return $j$
\EndFunction
\end{algorithmic}
\begin{algorithmic}
\Function{EstimateN}{$l$, $c$}  
\Comment $l$ can be computed from taget CV by eq. \eqref{eq:l from CV large N}
\State $X \gets 0$, $Y \gets 0$, $W_c \gets 0$ 
\For {$j=1 : l$}
 \State $W_c \gets \text{\sc getBlockSize($c$)}$
 \State $X \gets X + W_c$ \Comment accumulate $l$ block sizes
 \If {$W_c = c+1$} 
{$Y \gets Y+1$} \Comment count hits of memory limit $c$ 
 \EndIf
\EndFor
\State $\Return  \left\lfloor\frac{1}{1+\frac{0.27}{l}} \cdot \frac{2}{\pi}\cdot \left(X/l -\frac{2}{3}\right)^2\right\rfloor$, $Y$
\Comment see equation \eqref{eq:method 2 heuristic}
\EndFunction
\end{algorithmic}
\end{algorithm}

\section{Effects of limited memory on performace}
\label{sec:error}
In this section we consider the following question: at what values of memory limit $c$ the error due to block sizes that were clipped by the memory-restricted algorithm  
will remain small; say, one percent or less? 
We shall assume that $N$ is large and that sample size $l$ is big enough, so that the positive bias $\alpha \approx 0.27/l$ in the estimator can be neglected.

To answer this question we have done a series of experiments with pseudorandom sequence of symbols where the estimator in equation \eqref{eq:method 2 heuristic} used $W_c$ rather than $W$. This pseudorandom sequence was generated by Matlab's \texttt{randi} function. The alphabet sizes $N$ were $10^2$, $10^3$,\dots, $10^6$, and $c$ was set to $\lceil K\sqrt{N} \rceil$ with $K=2.5$, $2.6$, \dots, $3.0$, where $\lceil\cdot\rceil$ denotes the ceiling function: $\lceil x \rceil$ is the smallest integer $ \geq x$. This range of $c$ was chosen based on theoretical calculation that we will describe later.  

For each value of $N$ and $K$ an estimate $\hatN$ was computed by equation \eqref{eq:method 2 heuristic} after 109 block sizes have been collected and this operation was repeated twenty thousand times. 
\begin{table}
\caption{Empirical bias and CV as percentage points when block sizes are clipped at $c=\lceil K\sqrt{N} \rceil$; $l=109$.}
\label{table:censored estimator experiment}
%
{\tabulinesep=1.2mm
\begin{tabu}{|c|c|ccccc|}
\hline 
\multicolumn{1}{|c}{} &  & \multicolumn{5}{c|}{$N$}\tabularnewline
\hline 
$K$ &  & $100$ & $10^{3}$ & $10^{4}$ & $10^{5}$ & $10^{6}$\tabularnewline
\hline 
$2.7$ & bias \% & $-1.07$ & $-1.18$ & $-1.38$ & $-1.37$ & $-1.38$\tabularnewline
 & CV \% & $9.31$ & $9.52$ & $9.59$ & $9.69$ & $9.68$\tabularnewline
\hline 
$2.8$ & bias \% & $-0.83$ & $-0.88$ & $-1.02$ & $-0.99$ & $-1.01$\tabularnewline
 & CV \% & $9.35$ & $9.57$ & $9.70$ & $9.76$ & $9.75$\tabularnewline
\hline 
$2.9$ & bias \% & $-0.66$ & $-0.66$ & $-0.75$ & $-0.72$ & $-0.74$\tabularnewline
 & CV \% & $9.39$ & $9.61$ & $9.70$ & $9.82$ & $9.80$\tabularnewline
\hline 
$3.0$ & bias \% & $-0.54$ & $-0.48$ & $-0.55$ & $-0.53$ & $-0.53$\tabularnewline
 & CV \% & $9.42$ & $9.66$ & $9.74$ & $9.85$ & $9.84$\tabularnewline
\hline 
\end{tabu}
}
\end{table}
Empirical bias and coefficient of variation were computed from these results. (Recall that empirical bias is relative error between mean of the twenty thousand values of $\hatN$ and actual alphabet size $N$: $\text{bias} = (\text{mean}(\hatN) - N)/N$; empirical coefficient of variation CV is standard deviation of the twenty thousand values of $\hatN$ divided by their mean.)

In Table~\ref{table:censored estimator experiment} we have listed the bias and coefficient of variation as percentage points for a subset of measurements where $K$ is between $2.7$ and $3.0$. 

Notice, first, that because clipping data reduces its variablilty, the CV values of clipped estimator for any $N$ and $K$ in Table~\ref{table:censored estimator experiment} are less than the CV values of estimator for that $N$ and $K$ in Table ~\ref{table:bias and CV}.  

Second, the absolute value of bias decreases as $K$ grows from $2.7$ to $3.0$; and at $K=2.9$ it is less than one percent for all $N$ that we have tried. 

The conclusion from experimental data is that  storage capacity for $\lceil 2.9\sqrt{N} \rceil$ symbols should be enough to achieve one percent accuracy in the measurement. The time needed for doing one such measurement is at most $109 \cdot \lceil 2.9\sqrt{N} \rceil$. 
 
A theoretical calculation below leads to the same conclusion. 

Let us denote  by $\epsilon_1$ the systematic error of theoretical mean $E(W_c)$ relative to $E(W)$:
\begin{equation}
E(W_c)=E(W)\left(1-\epsilon_1\right).
\label{eq:E(W_c) error}
\end{equation}      

The fact that we are recording $W_c$, rather than $W$, together with $W_c \leq W$, introduces systematic error $\epsilon_2$ in our measurement of alphabet size $N$ that can be estimated by substituting $E(W)(1-\epsilon_1)$ in place of $E(W)$ in equation \eqref{eq:N from Q(N)}. After this substitution we discard the term $-2/3$, because in our scenario it can be reasonably assumed to be small compared to mean block size. The result is  
\begin{equation}
E(\hatN) = N(1-\epsilon_2) \approx \frac{2}{\pi} E(W)^2(1-\epsilon_1)^2 \approx N(1-2\epsilon_1 + \epsilon_1^2).
\label{eq:N error}
\end{equation}      
We see that 
\begin{equation}
\epsilon_2 \approx \epsilon_1(2-\epsilon_1).
\label{eq:epsilon2}
\end{equation} 

Next, we will derive an analytical expression for $\epsilon_1$ by applying the law of total expectation. For notational convenience we will denote the probability of the event $W>c$ with $p$ in that derivation. Firstly,
\begin{align*}
\begin{aligned}
E(W) = p \cdot E(W \mid W>c)
+ (1-p) \cdot E(W \mid W \leq c).
\end{aligned}
\end{align*}
By rearranging this equation, so that $E(W \mid W \leq c)$ is on the left hand side, we get
\begin{equation}
E(W \mid W \leq c) = \frac{E(W)-p \cdot E(W \mid W>c)}{1-p}.
\label{eq:E(W mid W leq c)}
\end{equation}
Secondly, 
\begin{align}
\begin{aligned}
E(W_c) = p \cdot (c+1)
+ (1-p) \cdot E(W \mid W \leq c),
\end{aligned}
\label{eq:E(W_c)}
\end{align}
Substituting the right hand side of \eqref{eq:E(W mid W leq c)} into \eqref{eq:E(W_c)} we obtain
\begin{align*}
\begin{aligned}
E(W_c) &= E(W)-p \cdot \left( E(W \mid W>c)-(c+1) \right) \\
&= E(W)\left( 1- p \cdot \frac{E(W \mid W>c)-(c+1)}{E(W)} \right).
\end{aligned}
\end{align*}

Therefore (cf.~equation \eqref{eq:E(W_c) error}), 
\begin{equation}
\epsilon_1= \Pr(W>c)\cdot \frac{E(W \mid W>c)-(c+1)}{E(W)}.
\label{eq:epsilon_1}
\end{equation} 

This enables to compute numerically the value of systematic error $\epsilon_1$ for a given $N$ and $c$:   
both $E(W \mid W>c)$ and $E(W)$ in the second term can be computed by equation \eqref{eq:E(W | W>c)} (the latter with setting $c=1$); and 
$\Pr(W>c)$  by equation \eqref{eq:Pr(W>k)}.  It is expedient to first compute logarithm of  $\Pr(W>c)$:
\begin{equation}  
\ln \left( \Pr(W>c)\right) = \sum_{k = 1}^{c-1} \ln (N-k) -(c-1) \ln (N), 
\label{eq: log(Pr(W>c))}
\end{equation}  
and then exponentiate the result.

When $c/N$ is small, the probability $\Pr(W>c) \approx \exp{(-c(c-1)/(2N))}$.%
\footnote{This approximation is obtained from $\ln(\Pr(W >c))$, using truncated Taylor series of the logarithm: $\ln(1-x) \approx -x$; see, e.g., Feller \cite{1968Feller}.} Substituting $K\sqrt{N}$ in place of $c$ in this approximation and simplifying, we get
\begin{equation}
\Pr(W>  K \sqrt{N}) \approx e^{\frac{-K^2}{2}} \qquad
\text{(if $K/\sqrt{N}$ is small).}
\label{eq: Pr(W>c) Feller}
\end{equation}   

In Table \ref{table:censored estimator theory} we have listed the calculated values of theoretical bias
$-\epsilon_2 \approx -\epsilon_1(2-\epsilon_1)$ as percentage points for $N=100$, $10^3$,\dots ,$10^7$ and $K=2.7$, $2.8$, $2.9$, $3.0$.
\begin{table}[h]
\caption{Theoretical bias as percentage points when block sizes are clipped at $c=\lceil K\sqrt{N} \rceil$.}
\label{table:censored estimator theory}
%
{\tabulinesep=1.2mm
\begin{tabu}{|c|cccccc|}
\hline 
\multicolumn{1}{|c}{} &  & \multicolumn{5}{c|}{$N$}\tabularnewline
\hline 
$K$ & $100$ & $10^{3}$ & $10^{4}$ & $10^{5}$ & $10^{6}$ & $10^7$\tabularnewline
\hline 
$2.7$ & $-0.76$ & $-1.10$ & $-1.31$ & $-1.36$ & $-1.37$ & $-1.38$\tabularnewline
\hline 
$2.8$ & $-0.53$ & $-0.81$ & $-0.96$ & $-1.00$ & $-1.01$ & $-1.02$\tabularnewline
\hline 
$2.9$ & $-0.37$ & $-0.59$ & $-0.70$ & $-0.72$ & $-0.74$ & $-0.74$\tabularnewline
\hline 
$3.0$ & $-0.25$ & $-0.43$ & $-0.50$ & $-0.53$ & $-0.54$ & $-0.54$\tabularnewline
\hline 
\end{tabu}
}
\end{table}

Based on this data it can be predicted that with $c \geq \lceil 2.9 \sqrt{N}\rceil$ the error of the estimator due to clipped block sizes will remain less than one percent. The experimental data summarized in Table \ref{table:censored estimator experiment} confirms this. 

In Table \ref{table:censored estimator theory and practice}  we have listed the difference between theoretical prediction of bias in Table~\ref{table:censored estimator theory}  and empirical bias in Table~\ref{table:censored estimator experiment}.
\begin{table}[h]
\caption{The difference between theoretical predictions in Table~\ref{table:censored estimator theory} and experimental data in Table~\ref{table:censored estimator experiment}.}
\label{table:censored estimator theory and practice}
%
{\tabulinesep=1.2mm
\begin{tabu}{|c|ccccc|}
\hline 
\multicolumn{1}{|c}{} &  & \multicolumn{4}{c|}{$N$}\tabularnewline
\hline 
$K$ & $100$ & $10^{3}$ & $10^{4}$ & $10^{5}$ & $10^{6}$ \tabularnewline
\hline 
$2.7$ & $0.35$ & $0.08$ & $0.07$ & $0.01$ & $0.01$ \tabularnewline
\hline 
$2.8$ & $0.30$ & $0.07$ & $0.06$ & $-0.01$ & $0.00$ \tabularnewline
\hline 
$2.9$ & $0.29$ & $0.07$ & $0.05$ & $0.00$ & $0.00$ \tabularnewline
\hline 
$3.0$ & $0.29$ & $0.05$ & $0.05$ & $0.00$ & $-0.01$ \tabularnewline
\hline 
\end{tabu}
}
\end{table}
At $N=100$, there is significant difference between theoretical bias and empirical bias -- theoretical calculation underestimates the bias; at the middle range of alphabet sizes, where $N=10^3$ and $N=10^4$ the difference is about four to six times smaller; and at the high range of alphabet sizes, where $N=10^5$ and $N=10^6$, the two are almost identical.

In summary, we have found that when number of blocks $l$ is set as $l=109$ in order to achieve ten percent precision in the measurement, and $c = \lceil 2.9 \sqrt{N} \rceil$, the underestimate of $N$ is less than one percent. With these settings the space and time used by the algorithm are at most $\lceil 2.9 \sqrt{N}\rceil $ and $109 \cdot \lceil 2.9\sqrt{N} \rceil $, respectively.

Furthermore, as we increase the memory limit $c$ beyond  $\lceil 2.9 \sqrt{N}\rceil$, the underestimate of $N$ decreases sharply. For example, when $c=\lceil 4.56 \sqrt{N}\rceil$, theoretical calculation shows that the underestimate of $N$ is in the order of $0.001$ percent. 

This phenomenon can be explained qualitatively by examining behavior of the two terms on the right hand side of  
equation \eqref{eq:epsilon_1}, as we increase memory limit $c$ from zero to $N$.

\xx{On the one hand, the value of the denominator $E(W)$ in the second term of \eqref{eq:epsilon_1} is independent of $c$: it is approximately $1.25 \sqrt{N}$ by equation \eqref{eq:Knuth simplified}.  When $c=0$, the value of $E(W \mid W>c)$ in the numerator of the second term of \eqref{eq:epsilon_1} also equals to $E(W) \approx 1.25 \sqrt {N}$; and in the other extreme, when $c=N$, the value of $E(W \mid W>c)$ is simply $N+1$. Therefore, the numerator of the second term of \eqref{eq:epsilon_1}, that is the difference $E(W \mid W>c) - (c+1)$, decreases from about $1.25 \sqrt {N}$ at $c=0$, to zero at $c=N$.}

When $c=0$, the value of $E(W \mid W>c)$ in the numerator of the second term of \eqref{eq:epsilon_1} equals to $E(W) \approx 1.25 \sqrt {N}$; and in the other extreme, when $c=N$, the value of $E(W \mid W>c)$ is $N+1$. Thus, the value of second term in equation \eqref{eq:epsilon_1} moves down from $1-1/E(W)$ at $c=0$, to zero at $c=N$.

The first term of \eqref{eq:epsilon_1}, that is the probability $\Pr(W >c)$, which equals to 100 percent when $c = 0$, takes a dramatic dive as we increase $c$ beyond $1.25 \sqrt{N}$.
%
For example, when $N=10^6$, it can be computed by equation \eqref{eq: Pr(W>c) Feller}   
 that  although 
$\Pr (W >  0.25 \sqrt{N})$ is $96.9$ percent, and
$\Pr (W > 1.25 \sqrt{N})$ is $45.8$ percent,   
already $\Pr (W > 3 \sqrt{N})$ is $1.1$ percent,
and $\Pr (W >  4.56 \sqrt{N})$ is $0.003$ percent only.

Moreover, the behavior of $\Pr(W > c)$, as we increase memory limit $c$ from zero to $N$, can be characterized by 
\begin{equation*}
\Pr(W > c) \leq e^{-\frac{1}{N}\cdot \frac{c(c-1)}{2}}  = e^{-\frac{c^2}{2N}}\cdot e^{\frac{c}{2N}} \leq e^{-\frac{c^2}{2N}}\cdot \sqrt{e},
\end{equation*}
where the first inequality is obtained from $\ln(\Pr(W >c))$ using the relation: $\ln(1-x) \leq -x$, which is valid for $0 \leq x < 1$; 
and the second, by noting that $\exp(c/(2N))$ attains its largest value $\sqrt{e} \approx 1.65$, at $c=N$.   
Thus, for instance, it can be computed that $\Pr(W > 10\sqrt{N}) \leq 2 \cdot 10^{-22} \cdot \sqrt{e}$.

Compared to the sharp decline of the first term of equation \eqref{eq:epsilon_1}, the decrease of the second term is rather gradual. Continuing the example of $N=10^6$, it can be computed that at $c = 0.25 \sqrt{N}$ the value of the second term is $82.7$ percent, at $c = 1.25 \sqrt{N}$ it is $46.1$ percent (at that point the first and the second term in equation \eqref{eq:epsilon_1} are almost equal), at $c = 3 \sqrt{N}$ it is $24.2$ percent, and at $c = 4.56 \sqrt{N}$ it is still $16.7$ percent. 

In conclusion, $\Pr(W>c)$ dominates equation \eqref{eq:epsilon_1} when the memory limit $c$ exceeds $1.25 \sqrt{N}$, and helps to drive down the size of the systematic error.

\section{Theoretical calculations}
\label{sec:theory}
The calculations in this section include five parts: 
\begin{itemize}
\item[] (\ref{sec:moments}) computation of the conditional moments $E(W^j \mid W > c)$;
\item[] (\ref{sec: Mean block size E(W)}) derivation of equation \eqref{eq:Knuth simplified} for mean block size $E(W)$;
\item[] (\ref{sec:E2 bias}) computation of the bias of the estimator; 
\item[] (\ref{sec:The variance of block size}) derivation of equation \eqref{eq:Var(W) simplified} 
for the variance $\Var (W)$ of block size;
\item[] (\ref{sec:coefficient of variation}) derivation of equations \eqref{eq:C^2 W result} and \eqref{eq:C^2 W)large N} for coefficient of variation. 
\end{itemize}

\subsection{Computation of the conditional moments $E(W^j \mid W > c)$}
\label{sec:moments}
We now turn to computing conditional moments $E(W^j \mid W > c)$, where $j$ is a positive integer, and $c =0$, 1,\dots, $N-1$. They are defined by  
\begin{equation*}
E(W^j \mid W > c)= \frac{1}{\Pr (W>c)} \cdot \sum_{k=c+1}^{N+1} \Pr (W=k) \cdot k^j.
\label{eq:moments definition}
\end{equation*}
Please note that since a block of input data starts and ends with a matching symbol, a block must include at least two symbols: $W \geq 2$. For that reason, $E(W^j \mid W> 0)$ and $E(W^j \mid W> 1)$ are the same as the unconditional moment $E(W^j)$. 

On the one hand, the probability distribution function of block size $W$ is well known, because it is needed in solving the Birthday Problem \cite{WikipediaBirthdayProblem}: 
\begin{equation}
\Pr(W \leq k)= 1- \frac{N^{\underline{k}}}{N^{k}} \quad \text{and} \quad \Pr(W>k)= \frac{N^{\underline{k}}}{N^{k}},
\label{eq:Pr(W>k)}
\end{equation}
where the symbol $N^{\underline{k}}$ denotes descending factorial
$N(N-1)\cdots (N-(k-1))$. 
The probability mass of event $W=k$ that we need to compute $E(W^j \mid W > c)$ is the difference between $\Pr(W>k-1)$ and $\Pr(W>k)$:
\begin{equation}
\Pr(W=k) =\frac{N^{\underline{k-1}}}{N^{k-1}}\cdot \frac{k-1}{N}.
\label{eq:Pr(W=k)}
\end{equation}

But on the other hand, direct computation of $E(W^j \mid W > c)$ from the definition can be numerically difficult for large alphabet sizes. We will therefore derive, via one-step analysis, a nested, computationally-efficient formula for $E(W^j \mid W > c)$. Below, the formula for $E(W^j)$ is derived first; it is then generalized to obtain $E(W^j \mid W > c)$.

Let us denote by $a_k$ the growth in exponential function $W^j$ when block size $W$ is incremented from $k$ to $k+1$:  
\begin{equation}
a_k = (k+1)^j -k^j.
\label{eq:a_k}
\end{equation}
(When $j=1$,  $a_k$ degenerates to one for all positive $k$. Also, $a_0=1$ independently of $j$.)

Suppose that we have observed $k$ symbols without finding a matching pair (that is, without seeing two identitcal symbols). Let us denote with $m_k$ the difference
 \begin{equation}
 m_k=E(W^j \mid W>k)-k^j.
 \label{eq:m_k general definition}
 \end{equation}
For example, in the simplest case of $j=1$, $m_k$ is the mean number of symbols that we will observe from that moment and until we find a matching pair; $m_k$ in this case is the difference between expected value of block sizes that are greater than $k$, and $k$.

Since there are only $N$ different symbols in the alphabet, the index $k$ of $m_k$ runs between 0 and $N$. If $k=N$, then the next, $(k+1)$st  symbol will surely match one of the previously seen; for this reason the value of $E(W^j \mid W>N)$ is $(N+1)^j$. From this and the above definition we know that $m_N$ is $(N+1)^j-N^j$. In the other limiting case of $k=0$, we have $m_0=E(W^j)$.

The next, $(k+1)$st symbol matches one of already observed symbols with probability $k/N$; and it does not match any of these symbols with probability $(N-k)/N$. In the first ``match'' alternative, $m_k=a_k$; in the second ``no match'' alternative, $m_k=a_k+m_{k+1}$. This defines a recursive relation between $m_{k}$ and $m_{k+1}$, for $k=0$, 1, 2, \dots, $N-1$:
\begin{equation}
m_k= \frac{k}{N}\cdot a_k + \frac{N-k}{N}\cdot \left(a_k+ m_{k+1}\right) = a_k + \frac{N-k}{N} \cdot m_{k+1}.
\label{eq:m_k}
\end{equation}

To solve this recursion it is expedient to first reverse its order (by formally replacing $k$ with $N-1-k$ in \eqref{eq:m_k}), so that the known term $m_N=a_N$ becomes first, and the term $m_0=E(W^j)$, which we wish to compute, becomes last: 
\begin{equation}
m_{N-k-1}= a_{N-k-1} + \frac{k+1}{N} \cdot m_{N-k}.
\label{eq:m_k reversed}
\end{equation}
To get a closed-form formula for $E(W^j)$ we start with $k=0$ and compute $m_{N-1}$ 
\[ m_{N-1} = a_{N-1} + \frac{1}{N}\cdot m_N = a_{N-1} + \frac{a_N}{N}.  
\]
Next we set $k=1$, and substitute what we have just computed into the right hand side of \eqref{eq:m_k reversed} in place of $m_{N-1}$. This operation is repeated for $k=2$, 3, and so on, until   at $k=N-1$ we get $m_0$. The resulting expression is
\begin{align}
\begin{aligned}
E(W^j) =
1 + & \left(a_1+\frac{N-1}{N}\left(a_2+ \frac{N-2}{N}\Bigg(a_3+ \right.\right. \cdots \\
 + & \left. \left(a_{N-2}+\frac{2}{N}\left( a_{N-1}+\frac{1}{N} a_N\right)\right)\cdots \right).
\label{eq:E(W^j)}
\end{aligned}
\end{align}

Finally, observe that by equation \eqref{eq:m_k general definition} the conditional moment $E(W^j \mid W> c)$, where $c=0$, 1, 2, \dots, $N-1$, can be computed in a similar manner, but starting the recursion \eqref{eq:m_k} from $c^j +m_c$ rather than from $m_0$. 
The resulting expression is  
\begin{align}
\begin{aligned}
E(W^j \mid W>c) = &
 c^j+\left(a_c+\frac{N-c}{N}\left(a_{c+1}+ \frac{N-(c+1)}{N}\Bigg(a_{c+2}+ \right.\right. \\
 &\cdots + \left. \left(a_{N-2}+\frac{2}{N}\left(a_{N-1}+\frac{a_N}{N}\right)\right)\cdots \right).
\label{eq:E(W^j | W>c)}
\end{aligned}
\end{align}
Setting $j=1$ in \eqref{eq:E(W^j | W>c)} we get the formula for computing the conditional expectation $E(W \mid W>c)$ that we need in section \ref{sec:error}:
\begin{align}
\begin{aligned}
E(W \mid W>c) = &
 c+\left(1+\frac{N-c}{N}\left(1+ \frac{N-(c+1)}{N}\Bigg(1+ \right.\right. \\
 &\cdots + \left. \left(1+\frac{2}{N}\left(1+\frac{1}{N}\right)\right)\cdots \right).
\label{eq:E(W | W>c)}
\end{aligned}
\end{align}
Let us illustrate this computation when $N=3$:
\begin{equation*}
E(W \mid W>c) = \begin{cases} 
  1+\left(1+ \frac{2}{3}\left(1+ \frac{1}{3}\right)\right) \approx 2.89, &\qquad c=1;\\
  2+\left(1+ \frac{1}{3}\right) \approx 3.33, &\qquad c=2;\\
  3+1=4, &\qquad c=3.
\end{cases}
\end{equation*}

Multiplying the terms in \eqref{eq:E(W^j)} gives   
\begin{equation}
E(W^j) = 1+ \sum_{k=1}^{N} \frac{N^{\underline{k}}}{N^k} a_k,
\label{eq:E(W^j) with Q}
\end{equation}
where the symbol $N^{\underline{k}}$ in the summand denotes descending factorial 
$N(N-1)\cdots (N-(k-1)).$


Equation \eqref{eq:E(W^j)} is better for numerical computations than equation \eqref{eq:E(W^j) with Q}. This is because the number of multiplications and divisions in equation \eqref{eq:E(W^j)} is about $N$ times smaller, and also because in evaluating \eqref{eq:E(W^j)} we do not need to multiply together many tiny, almost-zero numbers. 

Still, for $j=1$ there is a known asymptotic expansion of the second term of \eqref{eq:E(W^j) with Q}; and we will next use this result  to derive simpler formulas for mean and variance of block size $W$.

\subsection{Derivation of equation \eqref{eq:Knuth simplified} for mean block size $E(W)$}
\label{sec: Mean block size E(W)}
When $j=1$, equation \eqref{eq:E(W^j) with Q} reduces to
\begin{equation}
E(W) = 1+ Q(N),
\label{eq:E(W) Arnold}
\end{equation}
where $$Q(N) = \sum_{k=1}^{N} \frac{N^{\underline{k}}}{N^k}.$$

Asymptotic expansion of $Q(N)$ in descending powers of $N$ is given in Knuth's treatise \cite{1997KnuthVol1}. Plugging this expansion into \eqref{eq:E(W) Arnold} produces
\begin{align}
\begin{aligned}
E(W) &= \sqrt{\frac{\pi N}{2}} + \frac{2}{3} + \frac{1}{12}\sqrt{\frac{\pi}{2N}}
- \frac{4}{135 N} +  \frac{1}{288}\sqrt{\frac{\pi}{2N^3}} \\
&+ O(N^{-2}). 
\label{eq:Knuth}
\end{aligned}
\end{align}

The error from truncating the expansion in \eqref{eq:Knuth} to the first two terms on the right is about $1/\sqrt{N}$: that error is less than one for $N \geq 2$. We therefore truncate and obtain equation \eqref{eq:Knuth simplified}.

These results about the mean $E(W)$ are known. See Flajolet, Gardy and Thimonier \cite{1992Flajolet} and Sedgewick and Flajolet \cite{2013Sedgewick}.

\subsection{Derivation of equation \eqref{eq:Var(W) simplified} for the variance of block size}
\label{sec:The variance of block size} 

The function $a_k$ when $j=2$ is $(k+1)^2-k^2$, that is,  $a_k=2k+1$. In this case, equation \eqref{eq:E(W^j) with Q} reduces to
\begin{equation*} 
E(W^2) = 1 + \sum_{k=1}^{N}\frac{N^{\underline{k}}}{N^k}(2k+1)
=1+2 \sum_{k=1}^{N}\frac{N^{\underline{k}}}{N^k}\cdot k + Q(N).
\end{equation*} 

If we now rewrite the sum in the second term on the right hand side as a nested formula
\begin{align*}
\begin{aligned}
\sum_{k=1}^{N}\frac{N^{\underline{k}}}{N^k}\cdot k =
& \frac{N}{N}\left(1+\frac{N-1}{N}\left(2+ \frac{N-2}{N}\Bigg(3+ \right.\right. \cdots \\
 + & \left. \frac{3}{N}\left(N-2+\frac{2}{N}\left( N-1+\frac{1}{N} \cdot N\right)\right)\cdots \right),
\label{eq:Q(1,2,...,N) nested}
\end{aligned}
\end{align*}
then it can be noticed immediately that the contents of the innermost pair of brackets, 
$N-1+\frac{1}{N} \cdot N$, equal to $N$. Next we can compute the contents of the penultimate matching pair of brackets $N-2+\frac{2}{N} \cdot N$ and the result is again $N$, and so on, until in the end the contents of the outermost pair of brackets equal $N$ as well. Thus, the value of the whole sum is $\frac{N}{N} \cdot N = N$. 

All in all,  when $j=2$,
\begin{equation*}
E(W^2) = 1+ 2N +Q(N);
\label{eq:E(W^2) with Q}
\end{equation*} 
and since $E(W) = 1+Q(N)$, by equation \eqref{eq:E(W) Arnold}, we have
\begin{equation*}
E(W^2) = 2N +E(W).
\end{equation*} 

Finally, replacing  $E(W^2)$ by $2N+E(W)$ in definition of variance: $\Var(W) = E(W^2)-E(W)^2$,  results in equation \eqref{eq:Var(W) simplified}.

\subsection{The bias of the estimator}
\label{sec:E2 bias}

Formally expanding $g(\overline{W}_l)$  as a second-order Taylor polynomial around the mean $E(\overline{W}_l)$; applying expectation operator with respect to $\overline{W}_l$ to both sides; and then replacing in the resulting expression (i) $E(\overline{W}_l)$ with $E(W)$, and (ii) $\Var(\overline{W}_l)$ with $\Var(W)/l$, 
gives
\begin{equation}
E(\hatN) = E\left(g(\overline{W}_l)\right) \approx g(E(W)) + \frac{g''\left(E(W)\right)}{2} \cdot \frac{\Var \left(W\right)}{l}.
\label{eq:delta method++}
\end{equation}


The first term $g(E(W))$ on the right hand side of \eqref{eq:delta method++} can be replaced by $N$, because we have chosen $g(E(W))$ so that it is a rather good approximation to $N$. The second term in \eqref{eq:delta method++} is part of theoretical bias (systematic error) in our estimate of $N$; its absolute value decreases as the size $l$ of the sample increases. We denote the relative value of this bias by $\alpha$:
\begin{equation}
E(\hatN) = N(1+ \alpha), \quad \text{where $\alpha \approx \frac{1}{N}\cdot \frac{g''\left(E\left(X\right)\right)}{2}\cdot \frac{\Var \left(X\right)}{l}$}.
\label{eq:beta general}
\end{equation}

The function $g(W)$ is $\frac{2}{\pi}(W-\frac{2}{3})^2$. Its second derivative is the constant $4/\pi$. Recall that $\Var (W)= 2N - E(W)^2 + E(W)$. These and \eqref{eq:beta general} imply that
\begin{equation}
\alpha \approx \frac{1}{N}\cdot\frac{2}{\pi}\cdot\frac{2N - E(W)^2 + E(W)}{l}.
\label{eq:beta method 2}
\end{equation}

By equation \eqref{eq:var W, large N} below, variance of block size $W$ is about $0.43 N -1.67 \sqrt{N}$ when $N$ is large. Thus, the bias for large $N$ is
\begin{equation}
\alpha \approx \frac{1}{N}\cdot\frac{2}{\pi}\cdot\frac{0.43N -1.67 \sqrt{N}}{l} \approx \frac{0.27 - 1.67/\sqrt{N}}{l} \approx \frac{0.27}{l}.
\label{eq:beta method 2 large N}
\end{equation}

\xx{Recall that the function $g(W)$ in the estimator of Montalv\~ao et al.~\cite{2012Montalvao} is $a W^2 - b W + c$, with the coefficients $a=0.6366$, $b=0.8493$, and $c=0.1272$. The leading coefficient $a$ in this quadratic is the same as that of $\frac{2}{\pi}(W-\frac{2}{3})^2$ above. For that reason, both quadratics have the same second derivative, and it follows that  equations \eqref{eq:beta method 2} and \eqref{eq:beta method 2 large N} are valid for their estimator as well.
}

\subsection{Coefficient of Variation}
\label{sec:coefficient of variation}

Formally expanding $g(\overline{W}_l)$  as a first-order Taylor polynomial around the mean $E(\overline{W}_l)$; moving $g(E(\overline{W}_l)$ from the right to the left hand side; and applying expectation operator with respect to $\overline{W}_l$ to the square of both sides in the resulting equation, produces a fist-order approximation to variance of the estimator:
\begin{equation}
\Var(\hatN) \approx  \left[g'\left(E(\overline{W}_l)\right)\right]^2 \Var \left(\overline{W}_l\right).
\end{equation}
Replacing in that expression $E(\overline{W}_l)$ with $E(W)$, and $\Var(\overline{W}_l)$ with $\Var(W)/l$, 
gives
\begin{equation}
\Var(\hatN) \approx \left[g'\left(E(W)\right)\right]^2 \frac{\Var \left(W\right)}{l}.
\label{eq:var delta method++}
\end{equation}

This, together with first-order approximation $E(\hatN) \approx N$, gives  
the following approximation of coefficient of variation $\text{CV}=\sqrt{\Var(\hatN)}/E(\hatN)$:
\begin{equation}
\text{CV} \approx \frac{\left|g'\left(E(W)\right)\right|\cdot \sqrt{\frac{\Var(W)}{l}}} { N}.
\label{eq:CV}
\end{equation}

The first derivative of function $g(W)= \frac{2}{\pi}(W-\frac{2}{3})^2$ at point $E(W)$ is  $4/\pi (E(W)$ $-\frac{2}{3})$; its square is approximately  $8 N/\pi$ by equation \eqref{eq:N from Q(N)}.
Squaring both sides of \eqref{eq:CV} and substituting on the right-hand side (i) $8 N/\pi $ for the square of the derivative, and (ii) $2N - E(W)^2 + E(W)$ for the variance of $W$, results in equation \eqref{eq:C^2 W result}:
\begin{align*}
\begin{aligned}
\text{(CV)}^2 
&\approx \frac{1}{l}\cdot \frac{8}{\pi} \cdot \left(2-\frac{E(W)\cdot \left(E(W)-1\right)}{N}\right).
\end{aligned}
\end{align*}

Let us assume that $N$ is large. Then, firstly, $E(W)$ is much greater than one and by \eqref{eq:Var(W) simplified} the variance of $W$ is about $2N -E(W)^2$; and secondly,  
by \eqref{eq:Knuth simplified} $E(W)$ is about $\sqrt{\frac{\pi}{2} N} + \frac{2}{3}$, i.e. $E(W)^2 \approx \frac{\pi}{2}N + 2\cdot \frac{2}{3}\sqrt{\frac{\pi}{2}N}$. 
Thus, for large $N$ the variance of $W$ is about 
\begin{equation}
\Var(W) \approx 2N -\frac{\pi}{2}N - \frac{4}{3}\sqrt{\frac{\pi}{2}N} \approx 0.43 N -1.67 \sqrt{N} \qquad \text{(large $N$)}.
\label{eq:var W, large N}
\end{equation}

Squared coefficient of variation $\text{(CV)}^2$ of $\hatN$ for large alphabet sizes $N$ reduces to equation \eqref{eq:C^2 W)large N}:
\begin{align*}
\begin{aligned}
\text{(CV)}^2 &\approx  \frac{1.09  - \frac{4.25}{\sqrt{N}}}{l} \approx \frac{1.09}{l} \qquad \text{(large $N$)}.
\end{aligned}
\end{align*}

\section{Conclusion}
\label{sec:conclusion}
A variant of a simple algorithm for probabilistically estimating alphabet size $N$ from a stream of symbols output by a discrete uniform random source has been studied. It divides the stream of symbols online into adjacent blocks such that one block includes a single pair of identical symbols; and an estimate of $N$ is computed from the average size of $l$ of these blocks. 
That estimate's coefficient of variation  
(standard deviation divided by the mean)
is approximately $\sqrt{1.09/l}$ when $N$ is large.
 Increasing $l$ decreases coefficient of variation, which results in more precise estimate of $N$, but it also increases the average time (number of symbols) it takes to complete the measurement. When estimating large $N$ given a target coefficient of variation CV, $l$ should be set to $\lceil 1.09 /(\text{CV})^2 \rceil$.

\xx{The results of our study include  a more accurate formula for computing an estimate of $N$ from the average block size, and a formula that predicts the smallest number of blocks $l$ needed to achieve a given precision in the estimate of $N$.} 

We have analysed the effects of limited space (number of symbols that can be stored in computer memory) on accuracy of estimation, when algorithm that can store at most $c$ symbols replaces any block size greater than $c$ by $c+1$, and characterised the resulting underestimate of $N$.

It was found that when the number of observed blocks $l$ is set as $l=109$, in order to achieve ten percent precision in the measurement, $N=100$, $10^3$,\dots,$10^6$, and the space limit $c = \lceil 2.9 \sqrt{N} \rceil$, the underestimate of $N$ is less than one percent. With these settings, space and time used by the algorithm are at most $\lceil 2.9 \sqrt{N}\rceil $ and $109 \cdot \lceil 2.9\sqrt{N} \rceil $, respectively.
Furthermore, as we increase the space limit $c$ beyond $\lceil 2.9 \sqrt{N}\rceil$ the underestimate of $N$  decreases sharply. For example, when $c=\lceil 4.56 \sqrt{N}\rceil$, theoretical calculation shows that the underestimate of $N$ is about $0.001$ percent. 

In conclusion, the algorithm studied in this paper can quite accurately estimate large alphabet sizes $N$ using space and time of at most $\lceil K \sqrt{N} \rceil$ and $l\cdot \lceil K \sqrt{N} \rceil$, respectively, where the constant $K$ is in the order of 10.


\end{document}